# Designing for Meaningful Human Control in Military Human-Machine Teams

Jurriaan van Diggelen, Karel van den Bosch, Mark Neerincx, Marc Steen

TNO, the Netherlands

## Abstract

This chapter proposes methods for analysis, design, and evaluation of Meaningful Human Control (MHC) for defense technologies from the perspective of military human-machine teaming (HMT). Our approach is based on three principles. Firstly, MHC should be regarded as a core objective that guides *all phases* of analysis, design and evaluation. Secondly, MHC affects all parts of the socio-technical system, including humans, machines, AI's, interactions, and context. Lastly, MHC should be viewed as a property that spans longer periods of time, encompassing both prior and realtime control by multiple actors. To describe macrolevel design options for achieving MHC, we propose various Team Design Patterns. Furthermore, we present a case study, where we applied some of these methods to envision HMT, involving robots and soldiers in a search and rescue task in a military context.

## 1 Introduction

Current times witness an almost unbridled proliferation of military technology based on Information Technology, Robotics, and increasingly so on Artificial Intelligence (AI). Defense organizations worldwide have expressed their ambitions to use AI to gain a competitive edge by being faster, achieving information superiority, and a higher workforce capacity. Defense industry responds to this demand by developing a wide range of AI-based systems, such as unmanned autonomous systems, logistic support systems, decision support systems, operational and planning support.

Not surprisingly, the prospect of AI-based and autonomous military technology has led to a heated ethical debate. In 2015, over 3.000 AI experts have signed a letter that called for a ban on offensive autonomous weapons beyond meaningful human control (MHC) [17]. The fear is that this technology could lead to a dystopic future of dehumanized warfare. Within the last decade, many books, reports and articles have been written about this topic, discussing various possible future scenarios [26], the need for military competitiveness and a potential *AI arms race* [31], and legal and policy considerations concerning MHC in the military [12].

MHC requires humans to be closely engaged in any form of moral decision making (i.e. making decisions that affect human values). This human involvement may take a large variety of forms and is highly dependent on the context and the specific nature of the moral decision. This makes *a prohibition of all weapons that are not under MHC* subject to multiple interpretations. Furthermore, when not properly defined, such a ban could impose irresponsible restrictions on the military capabilities of those who would accept that law.

Besides regulating the abundance of new AI-based military systems coming on the market each year, a constructive approach is needed to operationalize AI responsibly. This includes a comprehensive design framework that addresses human, technological *and* contextual aspects with their mutual relationships. Rather than focusing on legal or technological solutions alone, MHC should be addressed as a process that must be designed in the socio-technical system at large. It should emerge from the interactions between humans and technology within a certain context over longer periods of time. Therefore, we adopt *the perspective of a human-machine team (HMT)*, allowing humans and machines to collaborate on a task to achieve a common objective [7]. This proceeds analogous to human team members that coordinate task with each other, adopt different roles matching their competencies, engage in collaborative planning, share situation awareness, and develop and learn over time. For MHC, it is particularly important to organize the team in such a way that the human remains in control over all moral decisions, i.e., the human adopts the role of moral agent.

The purpose of this chapter is to propose ways for analysis, design, and evaluation of MHC for defense technologies from the perspective of military human-machine teaming. In doing so, we will consider a broader range of technological systems than lethal autonomous weapon systems, where the MHC debate originated. Almost every AI-system in the defense domain poses moral challenges. For example, an AI-based situation awareness system that automatically classifies contacts into friendly and hostile, might well lead to use of force by human decision makers further down the chain. Moreover, besides *respecting human lives,* many non-lethal defense technologies still challenge other human values such as *privacy*, *safety* and *freedom*.

Characteristic of the defense domain is that operations are versatile, often highly unpredictable, and the ethical stakes can be extremely high. An analysis of MHC in this domain therefore always requires a deep understanding of the context. We propose to obtain this via stakeholder engagement and immersing them in a morally sensitive task to obtain a map of relevant values. Such analyses typically show that the temporal dimension of decision-making is particularly important. Human control is only meaningful when the human has sufficient time for decision making and to become aware of the context. The interaction between humans and machines must be designed in such a fashion that its supports planning in advance and replanning in case of unexpected events or unforeseen consequences.

Whereas this chapter focusses on defense technology, we believe that our analysis of how to design systems for MHC also provides valuable insights for achieving MHC and responsible AI in other domains, such as in healthcare and automotive.

The paper is organized as follows. The second section describes background on military ethics, military operations, and types of control, which form the foundation of our conception of MHC. The third section discusses a high level design framework for MHC in the defense domain, contributing to existing design guidelines such as those proposed by the defense innovation unit [10] or NATO[1]. In particular, we will propose various Team Design Patterns, as a way to implement MHC in a human-machine team. The fourth section presents a case study, where we applied some of these methods to envision HMT, involving robots and soldiers in a search and rescue task in a military context; in this case study, we collaborated with military end users. The conclusion summarizes our findings, and proposes future directions.

---

[1] An Artificial Intelligence Strategy for NATO, https://www.nato.int/docu/review/articles/2021/10/25/an-artificial-intelligence-strategy-for-nato/index.html

# 2 Background

To understand design considerations for MHC within military teams, we must first understand the current role of control and ethics within defense. The next subsections describe military morality, control, human-machine teams, and team development.

## 2.1 Morality within a military context

Whereas it may be difficult for some people to regard warfare as anything else than morally wrong, the ethics of warfare has been subject of legal and philosophical analysis since the ancient Greeks [35]. Perhaps the most famous tradition, so called *just war* theory, distinguishes between criteria that apply for starting a war (jus ad bellum) and criteria that apply during the conduct in war (jus in bello). For this paper, the latter field of study is most relevant. It states that, among other things, a distinction should be made between non-combatants and combatants (i.e. *principle of distinction*), and that the expected collateral damage resulting from an attack should be proportional to the military objective (i.e. *principle of proportionality*).

Whereas much has been written about these principles, and they have been codified in law in various ways (such as rules of engagement and international humanitarian law), applying these principles in military practice is never a self-evident effort [30]. For this reason, soldiers and military decision makers are trained in the application of these ethical principles during their basic military education. Whereas this type of training is crucial to effectuate ethics in military conduct, the moral dilemmas occurring in modern military conflicts remain difficult for even the most experienced soldiers. For example, consider a typical battlefield situation described by a US colonel who fought in Afghanistan [36]. *Taliban fighters are firing at American soldiers from the roof of a small apartment building. It is uncertain whether there are civilians in the building; possibly one or more families, possibly civilians held by the Taliban, possibly people finding shelter from the war. The US soldiers distinguish three options to act: (a) destroying the building by opening artillery fire or calling in an air strike, (b) approaching or entering the building and aim directly at the militants, and (c) withdraw troops and surrender the place to the enemy.* Each option has different consequences for own safety, safety of civilians, and reducing Taliban threat. Within this context characterized by uncertainty and conflicting values, the principles of distinction and proportionality offer little guidance how to navigate this dilemma.

To be able to say anything sensible about morality in military conduct, a thorough understanding of context is required. Morality and military context have an unsteady relation, where small changes in context can have large implications for the moral assessment. Some important factors are whether the activity is defensive or offensive, lethal or non-lethal, secret or not, whether or not civilians are present, the degree of uncertainty, etc. In modern military practice, the context is only getting more complex, continuously posing new challenges for war ethics. One challenge is how to deal with enemies that do not adhere to the same moral standards. For example, how to respond responsibly to an enemy that has committed itself to unbridled AI-based robotic warfare. Another challenge is that the enemy might use our own moral standards against us. Examples are using human shields, hiding weapons or combatants in protected areas such as hospitals and houses of worship. Another adversary tactic which is increasingly used is *information warfare*: spreading misinformation or propaganda to influence public opinion and win the political war. For example, the enemy may share manipulated videos on social media with the intent to "demonstrate" immoral behavior of our troops.

To summarize, like in other morally sensitive domains (such as healthcare and automotive) the **moral complexity of the military context** is characterized by opposing values, uncertainty, and evolving public opinion. However, a few factors make the military context unique and are important for designing responsible military AI. These factors of the environment, tasks, and actors are crucial to understand morality in the military domain. They include at least: adversary tactics, uncertainties in the operating environment, presence of civilians, defensive/offensive, lethal/non-lethal, presence of human/non-human targets, public opinion.

## 2.2 Moral decision making within military Command and Control

Command and control (C2) is the exercise of authority and direction by a properly designated individual over assigned resources in the accomplishment of the mission [8]. For centuries, it has been recognized that C2 is crucial to ensure responsibility, accountability, agility, and gaining military superiority. To understand morality in military C2, we must first recognize that there is no single decision maker. In modern warfare, C2 is a distributed responsibility [1].

As argued in [13], the chain of decisions that is relevant in the lethal autonomous weapons debate can be understood in 24 stages: from the political decision to go to war, to developing strategy, to selecting a target, to executing an attack. These stages have been metaphorically compared to an iceberg, as only the last stages of the process are visible and result in physical manifestations of harm. In cognitive engineering, the phenomenon is widely recognized and sometimes referred to as the *blunt and pointy end of the spear* [15]: only the pointy end hits reality, but it is supported by the blunt end of the spear. No matter which metaphor is used, the lesson to be drawn is that there is no single moral decision maker.

We can define a moral decision as a decision that affects human values. Human values that are relevant in defense are physical integrity (which is clearly violated in the case of weaponized attacks), but also values that may be less obvious, such as privacy, freedom, safety, comradery, patriotism. Moral decisions are made throughout the entire chain of command. Not only by the political leaders who decide to go to war, and not only by the people that are close to executing the attack or other activity that results in harm. Each of them has their own moral considerations using the information that is available to them, at their level of command, at that moment in time.

Adopting a different role in the moral decision-making process has consequences for attributing moral responsibility [18]. Sometimes it may be difficult, if not impossible, to attribute moral responsibility to a single person. This phenomenon is also known as the *problem of many hands* [22]. Over the centuries, military organizations, lawyers and ethicists have developed ways to work with it, and to ensure moral responsibility and accountability. The inclusion of AI in these processes requires us to further develop these solutions.

Summarizing, **moral decisions** are decisions that affect human values, where
- the effect of the decision may be brought about directly or further down the decision chain.
- the nature of the decision is highly context-dependent.
- human values include physical integrity, privacy, safety, equality, etc.
- human values may regard the decision maker's own values, or those of others.
- the decision could also be to refrain from action or to delegate to another decision maker.

## 2.3 Military human-machine teams and meaningful human control

Technological artifacts have always played an important role in the military. Relatively recent in history, robots and computerized systems have been added to the standard military's arsenal. Most of these systems (with a few potential exceptions) qualify as tools: they do exactly what they are instructed to do and do not possess any AI, nor do they have autonomy over their own behavior. For example, the weaponized drones that were widely used by the US in response to the 9/11 attack were all tele-operated by humans and did not undertake any activity unless explicitly instructed by a remote human operator. Although these systems certainly raised ethical concerns [24], and posed complex control problems for engineers, these are not the systems that sparked the debate on MHC. This is because the moral decisions (like weapon release) are made by humans, and the robot only *carries out* those decisions. MHC becomes a relevant issue when the robot is powered with AI and possesses a certain degree of autonomy over its decision-making, allowing it to act without being fully controlled by a human.

First of all, we should ask ourselves the question why the military would want such systems in the first place. There are, at least, three reasons. Firstly, the remote connectivity may be insufficient to allow teleoperation by a human. This can be due to environmental factors (e.g. indoor, in a tunnel, etc.), or because the enemy is deliberately interfering with the radio signals. Secondly, the tempo may be too high to allow human teleoperation (e.g. in cyber operations operating at machine speed, or in providing air-defense for hypersonic missiles). Thirdly, the number of robots and AI-based actors may be so large that it would require too many human tele-operators (e.g. as is the case with robot swarms [32]).

Given that there is undeniably a push towards more autonomous systems in the military domain, it is important that – no matter how advanced these systems become - there will always be some interaction with humans at some stage, e.g., in preparation, issuing commands, or during mission execution. Therefore, most defense organizations embrace the paradigm of human-machine teaming [27], in which one or more humans and one or more machines work together for a longer period of time to achieve a certain objective.

The key to achieving MHC in a human-machine team is to establish a proper task division among the human and machine members of the team, and to allow this to change dynamically if the circumstances demand this [34]. Within such human-machine teams, the human must play the role of *moral agent* (i.e. make all moral decisions), and the machine must be prevented to make any moral decisions [29]. This still allows the machine to autonomously perform other non-morally sensitive tasks (such as locomotion). Whereas this may seem like a simple solution for achieving MHC, two things make it complicated.

Firstly, allocating moral tasks to the human must not result in imposing an impossible task on the human. This is the case when the human has insufficient time or is insufficiently informed about the decision it must make. For example, an autonomous car on the highway should not unexpectedly transfer full control to the human to deal with a difficult traffic situation. This would result in a *moral crumble zone* [14]*,* where the human becomes scapegoat for faults in system design. The human never was in control due to a wrongly designed system which did not provide sufficient time and situation awareness for the human to intervene.

Secondly, there's the issue what it means not to be engaged in moral decision making. This is a difficult question, as the time of decision making does not necessarily coincide with the time the decision is carried out. For example, we probably would not say that a guided torpedo engages in moral decision making when it hits its target. The moral decision has been premade by humans at the time it left the mothership, and the torpedo only carries out the decision. But what if the target is less precisely described, such as with an intelligent naval mine that is programmed to explode when it recognizes a predefined acoustic signature of a ship? Or what about an unmanned surface vessel that is programmed to neutralize a contact when it classifies it as a threat? These examples show that there's a spectrum between a machine that carries out a moral decision, and a machine that autonomously makes a moral decision. The boundary is not clearly definable and depends on expert judgement whether the machine actually has some freedom in interpreting the commands issued by the human or not.

Summarizing, **Meaningful human control** implies that humans are enabled to make all moral decisions, and machines do not make any moral decisions, where
- *enabling* humans to make decisions includes providing the right conditions, such as providing sufficient time, situation awareness, information, etc.
- machines not *making* any moral decisions does not mean they are prevented from *carrying out* moral decisions that have been premade by humans.
- preventing machines from *making* any moral decision precludes any autonomous machine behavior that influences the decision's outcome.

## 2.4 Temporal aspects of meaningful human control

In Section 2.2, we argued that control in military teams is executed at various stages at different levels of abstraction, at different times, by different actors. Examples are by a military commander doing mission planning, by the briefing officer that provides a mission brief, or by soldiers that execute a mission. Within this chain of commands, each command is used to adjust to newly available context information by making a previously issued command more precise, or by redirecting the previously planned course of action.

When humans are controlling AI-based systems, the situation is not different. For MHC this means that control is executed at different moments in time. Using *prior control*, the human controls the system before the moral problem arises. Using *realtime control*, the human controls the system at the same time the moral problem arises, e.g. by temporarily switching to teleoperation mode.

The advantage of realtime control is that all context information is available (provided that this is sufficiently communicated to the human). This could lead to better-informed decision-making. A disadvantage is that there may be insufficient time to engage in extensive moral deliberation at the moment the moral dilemma occurs. Prior control, on the other hand, allows more time for moral deliberation. However, the further the moment of control is away from the actual moral dilemma, the less accurate context information is available, leading to less-informed moral decisions. Typically, the optimal solution will support prior control to pull forward moral decisions as much as possible, and will support realtime control to make last-minute adjustments to the plan.

Another temporal dimension of MHC is that humans will become more familiar with the systems they are interacting with, being able to better predict the system responses. Over time, this could lead to better MHC as the team develops.

Summarizing, **meaningful human control** may be executed by multiple actors at different moments. We therefore need to think of MHC as an emergent property that emerges from the interactions between multiple humans and technology over a longer period of time. Using **prior control**, moral decisions are made before the moment the problem occurs. Using **realtime control**, moral decisions are made at the same time, or shortly before the problem occurs

In the next section, we describe how these different solutions can be applied to design a human-machine team under MHC.

# 3 High-level design framework for Meaningful Human Control

In the previous sections, we have argued that MHC is a dynamic and complex property which has socio-technical, moral, and temporal dimensions. Consequently, designing MHC is a complex process as well, and cannot be captured in a simple checklist, or waterfall design method. To achieve and maintain MHC over a human-agent team, careful considerations are needed during the iterative phases of Analysis, Design, and Evaluation of the system. Some relevant considerations are depicted in the Figure below.

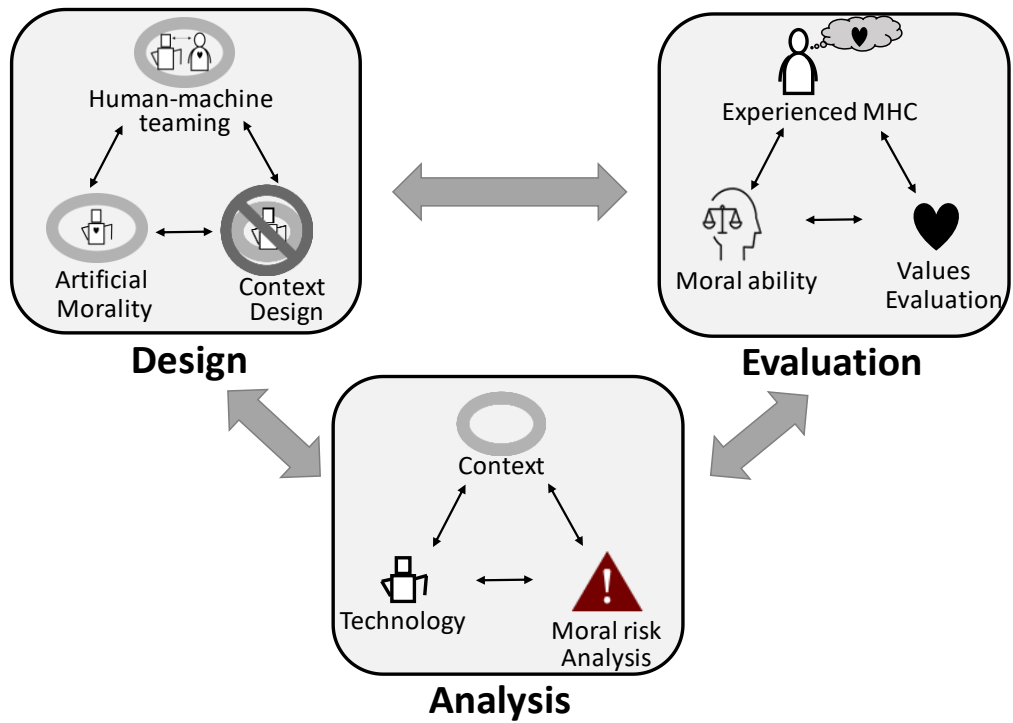

*Figure 1: High-level design framework for MHC in HMT's.*

This section discusses the main considerations for each phase and provides pointers for further reading. We will zoom in on design solutions for Human-Machine Teaming and provide some tools for addressing the temporal aspects of MHC.

## 3.1 Analysis

The following subsection discuss the analysis of technology, context, and moral risks.

### 3.1.1 Technology

To design for MHC, it is important to understand some aspects of the technology that forms part of the larger socio-technical system. NATO distinguishes three types of artificially intelligent technological systems [23]. The first type of AI is based on algorithmic search and instructions that are preprogrammed by humans. Systems of this type typically behave deterministically, which facilitates control by human operators. Examples of military systems of this type are: guided torpedoes and military planning software. The second type of AI is based on machine learning (or deep learning). These systems are capable of learning from historical data that enables them to make decisions on previously unseen cases. The human may exert control on a self-learning system by selecting the training examples (a form of prior control). For example, a system-designer may choose particular image classifiers to influence how a system learns (e.g., for a system that autonomously recognizes objects in camera streams). The third type of AI uses computational models of its social environment to improve understanding and interaction with it. These systems can cope with uncertainty and the complexities of an open world. They behave adaptively, taking the situational context into account. As a result, these systems are less predictable for humans, and more difficult to exert control over. The first operational applications of such systems include self-driving logistic vehicles, and highly autonomous unmanned aerial vehicles (UAV's).

One may argue that a fourth category could be added: the hypothetical case of artificial general intelligence (AGI), or artificial super intelligence (ASI). These systems can be expected to create control problems of an unprecedented nature [6]. However, as it is unclear whether, when, and in which form, these forms of AI will appear [43], they fall beyond the scope of this paper.

Depending on the type of AI technologies involved, we may ask the questions: What sensor data are available to the technology? How does it use sensor data to develop situational understanding? How does it develop plans? What goals does it pursue? How does it adapt and learn over time?

### 3.1.2 Context and task Analysis

The analysis of the military context serves to comprehend what moral issues may arise during operation (see Section 2.1). Such understanding is crucial for designing for MHC. The importance of studying real-world context in system design has been emphasized by the cognitive engineering community for decades. Fortunately, many useful tools exist to conduct such analyses. For example, cognitive work analysis [20] offers useful tools for systematically identifying stakeholders, tools, procedures, environmental constraints, etc. Typically, a context analysis is conducted in collaboration with domain experts, and results in several models of decision making and scenarios. Relevant questions are: Who makes moral decisions at which times? Which situation awareness do they need for this? Which information do they have available? Under what time pressure do they act? Crucially, these scenarios do not only capture the nominal cases (when everything goes as expected) but they also provide what-if scenarios to describe things that could go wrong. A wicked problem concerns the analysis of the mutual adaptative processes between humans and machine-learning agents that evolve over time.

### 3.1.3 Moral risk analysis

The developed scenarios can then be used for moral risk analysis. This allows us to better understand where and when MHC is needed. Several practices from value-sensitive design can be applied here [33]. A list of relevant values must be formulated, and put into context of the scenarios (for the nominal flow of events, but also for the what-if scenarios). This allows us to identify critical decision points where value tensions take place. For example, when a robot decides to inspect a house for potential terrorist activities, it may choose to sacrifice one value (e.g. privacy of its residents) for the sake of another value (e.g. safety of the community). MHC demands that these decisions that affect human values are made by humans (as argued in Section 2). Typically, this analysis is done with diverse stakeholders (e.g. military personnel, ethicists, and representatives of civilians). The stakeholders must be sufficiently immersed in the scenario to be able to appreciate these tensions. This could be done via narrative immersion (i.e. using a cognitive walkthrough, or by co-developing the scenario), but also using VR techniques.

## 3.2 Evaluation

Much of the academic discourse on MHC is about analysis. To make progress in the field, we must also gain practical experience with different concepts. The mantra *"fail fast, fail often"* captures the idea of gaining experience while trying out things and starting small. However, when it comes to moral risks, we must be careful not to fail in the real world and should identify potential failures before the system leaves the lab. This requires iterative lab-testing of different HMT designs against relevant measures for MHC. During these lab experiments, the following measures can be used (see also [34],[39]):

1) **Experienced MHC** corresponds to the subjective experience of control by humans in the HMT when collaborating with the AI-system. This can, for example, be measured using questionnaires and interviews.
2) **Moral ability** measures whether the decision maker had sufficient expertise, time and situation awareness to make the right decisions and understand their moral complications. For actors involved in realtime control, this requires that their operator control units provide them with sufficient information. For actors involved in prior control, this requires that they can predict the moral context with sufficient accuracy.
3) **Values evaluation** measures whether the team behavior corresponds with the human's moral values. In contrast to documented ethical guidelines, a human's moral values is not directly accessible. A possibility to assess the human's moral values is by asking in hindsight whether the team members found their team decisions ethically appropriate.

## 3.3 Design solutions

This section describes solutions for designing systems that support human-agent teams to operate under MHC. We propose to categorize these solutions as follows: (1) banning the use of autonomous technology in a particular context; (2) better teaming by enhancing Human-machine collaboration; (3) developing capabilities of autonomous moral decision making by the AI system. When and how to apply these solutions is discussed in the following subsections.

### 3.3.1 Ban use in certain contexts

Just like chemical weapons are banned by the Geneva Convention, one might argue that any <u>L</u>ethal <u>A</u>utonomous <u>W</u>eapon <u>S</u>ystem (LAWS) that is not under MHC, should be banned too. In fact, this solution is advocated by the "stop killer robots" campaign[2], and by the peace organization PAX. However, unlike detecting illegal chemical substances in weapons, detecting the presence or absence of MHC is far from straightforward. As argued, MHC is not a property of the technology alone, but is a relational property between humans, technology, and context. Banning a technology altogether also excludes its potential benefits, which could be immoral in itself. For example, is it moral to expose soldiers to a direct line of fire because the deployment of a robot was dismissed altogether based on not meeting the requirement for MHC under all possible operational conditions?

As a solution to harvest the potential of intelligent autonomous technology whenever possible, and to exclude its use whenever moral issues arise that necessarily demands human control, we advocate to restrict autonomy of technology in certain pre-specified contexts. For example, the use of weaponized land robots may be restricted to open battle fields, and to ban their use in urban environments, where the presence of civilians is more likely. This policy corresponds to what is already operational practice in some military context. Consider, for example, the naval anti-missile system of defense company Thales: *goalkeeper*[3]. This system is able to fire autonomously. The autonomous mode of this system is used in a well-defined context only: in open waters, against high-speed incoming missiles, where falling debris is less harmful and targets are clearly identifiable.

---
[2] https://www.stopkillerrobots.org/

From a design perspective, the context should be regarded as part of the *moral operational design domain* (MODD) [29]: the set of conditions in which the technological system, given the way it is designed, will operate in way that is considered as morally acceptable. System design and operational use must ensure that the system adheres to this domain. For some systems, it may be that the MODD is empty: no context or way of using it would be morally acceptable, which boils down to banning the system altogether. This is true for systems capable of doing much harm that are too unpredictable in their behavior to be kept under MHC.

### 3.3.2   Improve human-machine teaming

Improving realtime collaboration is often interpreted as *human-in-the-loop, human-on-the-loop,* or *supervisory control* [28]. Whereas this may help to achieve MHC in a very direct way (particularly for the deterministic type of AI systems mentioned in Section 3.1.1), it burdens the human with the dull and time-consuming task of acting as a failsafe. Supervisory control is not sufficient for interacting with more advanced AI-based systems. By instead designing the AI-based systems as a teammate, the AI is able to manage its collaboration with humans more adaptively and actively, analogous to the way human teammates collaborate.

Supporting humans and AI systems as teammates requires a carefully designed system [19]. Critical team functions include sharing situation awareness, understanding each other's role, managing interdependencies, aligning goals and plans, etc. [21]. This chapter focusses on the temporal dimensions within an HMT, crucial for achieving MHC.

Teams are dynamic entities, meaning that the way tasks are distributed over team members changes over time and in response to situational task requirements. To design these dynamics, team design patterns (TDP's) have been proposed as reusable solutions to recurring problems regarding human-machine collaboration [11]. TDP's have also been developed for teams that operate in moral contexts [34]. How various patterns can be applied to achieve MHC is described below.

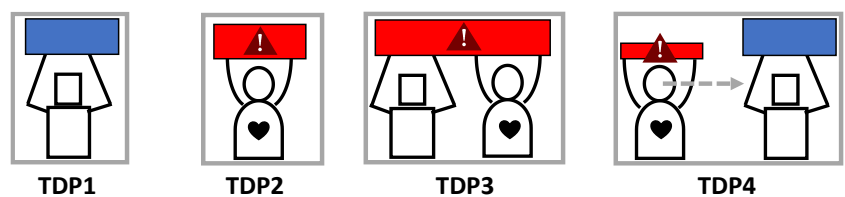

*Figure 2: Some basic Team Design Patterns involving moral tasks*

Figure 2 illustrates four basic team design patterns for distributing moral tasks (depicted as red blocks with an exclamation sign) and non-moral tasks (blue blocks) over humans (round figures) and machines (angular figures). TDP1 describes the pattern of a machine that performs a non-moral task. TDP2 show a human performing a moral task, thereby acting as a moral agent. The heart in the figure indicates that it can also process non-rational factors into its performance, like e.g. emotions and intuition.

---

[3] https://www.thalesgroup.com/en/goalkeeper-close-weapon-system

Machines without the heart symbol cannot perform moral tasks like humans, as they lack the necessary evolutionary background and capability to appreciate the emotional impact of possible decisions [42]. But they can, however, perform a moral task in close collaboration with humans (as shown in TDP3), as a joint activity [19]. An example would be a military decision support system with the AI supporting the human using a course of action (COA) analysis, risk prediction, collateral damage assessment, etc. Explainable AI can be applied to prevent the human from trusting the system too much or too little (e.g. blindly following the AI's advice, or ignoring all advice).

TDP4 shows yet another type of collaboration where the human monitors whether the machine refrains from morally sensitive tasks. This type of teamwork particularly makes sense when it is also specified what to do if moral sensitivity is detected, which can be specified using *composed patterns* as depicted in Figure 3.

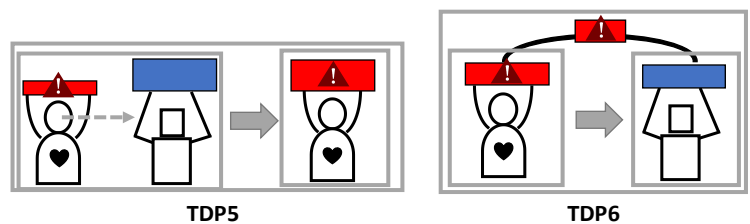

*Figure 3: Some composed Team Design Patterns that can be used to achieve MHC*

TDP5 shows a combination of TDP4 and TDP2, with a thick grey arrow between them depicting a temporal relation. This pattern describes the machine offloading the human by doing non-moral tasks, but when the human detects that the situation involves moral decision making, the human takes over. For example, a drone conducts area surveillance autonomously at low altitude, as stated in the military operation plan. During the operation, unexpected things may turn up such civilians being present in the area. When the human detects that the drone observes civilians, moral issues come into play. The human is to either give priority to the value of thorough area inspection, or the value of preserving civilian's privacy. Depending upon the human's value consideration, the human maintains a low elevation of the drone, or it directs the drone to fly at higher altitude. The patterns should also address the conditions and procedures for transferring tasks between team members. For example, the human must be at the ready for take-over considering, for example, availability and situation awareness [11].

Another pattern for human-agent team collaboration prevents machines to engage in moral decision making is depicted in TDP6. In this TDP, the original task is decomposed into a morally sensitive part (which is performed by the human in the preparation phase), and a non-morally sensitive part, which the machine can do during the operation. For example, before commencing the drone surveillance mission, the human decides that the drone is prohibited to fly over civilians at low altitude, and the drone adheres to that during its mission. The TDP6 is an example of achieving MHC by means of *prior control*. TDP7 and TDP8 are variants (see Figure 4.).

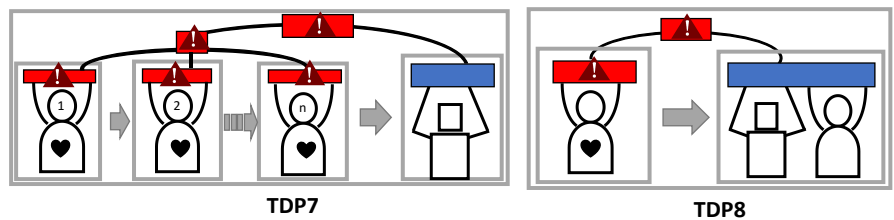

*Figure 4: Alternative patterns to TDP6 for prior control.*

TDP7 shows a variant where human prior control is performed by multiple actors (see Section 2.4). The variant TDP8 combines prior control with control at the moment a moral decision is actually carried out. This pattern supports the human to execute control until the very last moment, taking all actual situational circumstances into account before a final decision is being made. This pattern may help preventing humans to sustain the *sniper's syndrom*, an often-experienced phenomenon by UAV pilots who suffer emotionally from killing people at a long range [5].

It should be noted that TDPs describe the interactions within a team at an abstract level, and need to be applied in the specific context by the team members. The patterns thus support achieving MHC, but do not guarantee MHC by themselves. The conditions for achieving MHC must be implemented by the team or by the higher organization. Such conditions may be: ensuring that a task transition does not capture the team-members by surprise, or ensuring that the human is sufficiently aware of the context when engaging in prior control tasks. What context-dependent aspects must be arranged to ensure MHC is insufficiently understood and is therefore part of the research agenda for the coming decade [21].

### 3.3.3   Artificial morality

A profoundly different approach to achieving morally acceptable machine behavior is to use so-called Artificial Moral Agents (AMA's) [9]. AMA's use a computational model of morality to make moral decisions autonomously. In our symbology, this is depicted as a machine with a heart, which can therefore also perform moral tasks (as in TDP9). For example, the surveillance drone, when implemented as an AMA, would know how high it should fly in certain situations, without requiring a prior briefing by a human as in TDP6.

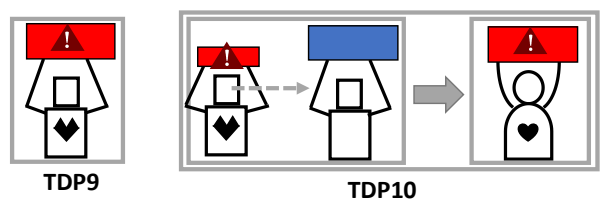

*Figure 5: Artificial Moral Agents*

Much controversy exists around artificial moral agents [37], and it is generally agreed that the required technology is still in its infancy [9]. However, in many cases, a light form of moral agency can also be helpful. For example, TDP10 is a variant of TDP5, where the machine only uses moral reasoning to *recognize* moral sensitivity, but leaves it to the human to *make* moral decisions.

Some would argue that AMA's are incompatible with the idea of MHC [3], as a machine is making moral decisions. Lacking a conscious experience of emotions such as grieve, fear and regret, the computational moral model may only be a shallow simulation of morality (hence the angular depiction of the heart in TDP9).

An important aspect is how the AMA acquires its computational moral model [2]. If all values captured in the artificial moral model stem from humans, for example through a process of value alignment [25], we might argue that the decisions made by these machines are (indirectly) human-controlled. In fact, when highly capable machines act in complex task settings, this type of prior control might be the only one feasible. Nevertheless, even more than for the previous approaches, such systems must be evaluated thoroughly.

# 4   Designing for MHC in Practice

In the previous sections, we discussed various ways to design for MHC within an HMT. The literature reports several recommendations, but in practice, there are no fixed rules about which methods to use or in which order. This depends, for example, on the availability and readiness of the technology, the intended deployment in the work organization, the availability and willingness of stakeholders to participate in the development, the size of the project (both in terms of budget and lead time), the project team's knowledge and experiences with such technology and related MHC-aspects, etcetera. Designing for MHC in practice requires a pragmatic attitude and creativeness to come up with new methods or ways of applying them when the project demands this.

As an illustration, this section presents the activities we undertook within a representative explorative project in the military domain. Our customer, the Dutch Ministry of Defense, has expressed the ambition to deploy human-machine teaming, and autonomous and AI-based robots to meet future challenges and threats [40]. At the same time, the Defense organization feels committed to responsible innovation, and to ensuring and maintaining MHC ([40], p.20). In collaboration with military domain experts, we explored the needs and opportunities for MHC over medium-sized robots within the very early phases of innovation. The following sections briefly describe our findings within the nine categories of the design framework, as presented in Figure 1.

## 4.1   Analysis

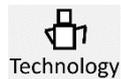
In the project, we adopted a dog-shaped robot as a representative example of the envisioned autonomous technology. The robot is comparable to Boston Dynamics' SPOT. The robot is equipped with cameras, heat sensors, microphones and speakers (for playing pre-recorded or real-time sounds). Furthermore, the robot is capable of autonomous navigation and mobility [4] and image recognition. The robot was deliberately *not* equipped with weapons, as the Dutch MoD is not envisioning the deployment of such systems. But the deployment of non-armed robots can be controversial as well. For example, a non-armed dog-shaped robot for use in police operations has raised ethical concerns, i.e. that the robot might become weaponized in the future, and that the robot would be disproportionately deployed in already marginalized and at-risk populations [38]. An ethical analysis of such robots in a military context is therefore appropriate. For our analysis we assume that the robot has the kind of autonomous capabilities that can be expected in about eight years from now. For example, it is believed that the robot will not require continuous human supervision for its functioning. However, at the same time, it is not believed that the robot will be able to cope with the world's unpredictability as humans can.

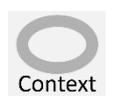
To provide a context of a possible military deployment of such a human-robot team, we developed a scenario of an asymmetric conflict. The scenario describes a mission, based on a (fictive) resolution of the United Nations Security Council to restore peace and security in the fictious country Maresland, which suffers from violence by a separatist and terrorist group. The scenario starts with an explosion in a residential flat in a middle-sized town. The explosion may have been caused by an accident or by a terrorist attack, this is yet unknown. One military team is tasked with searching and rescuing victims. Another military team is tasked with, covertly, collecting intelligence on possible perpetrators, and preventing a potential second attack. The operations are risky, under time pressure, and fraught with uncertainty.

The scenario involves several moral risks that could be identified upfront: e.g., harming safety of civilians (in case of a second attack; or when victims are not rescued in time); invading people's privacy while collecting intelligence; putting own troops at risk.

## 4.2 Design

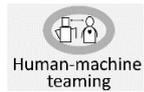 We assumed that both human-agent teams consist of three dismounted soldiers and five co-located robots. In this first iteration of the design we did not make any additional assumptions about mission phases (such as training, mission preparation, and evaluation), nor did we commit to any specific Team Design Patterns for the teams yet. The generic design choices were incorporated in the scenario, leading to a first *design scenario* [41].

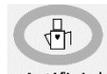 One of the objectives of this evaluation was to gain insight as to what moral issues might come forward when deploying intelligent technology in human-agent teams on such military missions. For that reason, we did not equip agents with artificial morality. This ensures to bring potential consequences of moral dilemmas firmly to the surface. For the same reason, we did not apply Team Design Patterns that rely on artificial morality (such as TDP9 and TDP10).

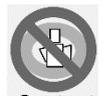 We extended the design scenario with a number of contexts that were considered interesting for investigation: (1) the robot acting within line of sight of the human; (2) the robot acting beyond sight of the human; (3) the robot performing a house search semi-autonomously; (4) the robot assisting in triage of victims; (5) the robot acting in a communication-denied environment (due to loss of network connection).

## 4.3 Evaluation

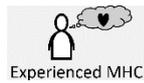 To investigate whether humans would experience MHC over missions carried out by such teams, eight soldiers were invited to our Human-Machine teaming lab for an immersive experience of the mission. The experiment uses a custom-built virtual reality environment, which featured a treadmill, so participants could 'walk' through the constructed scenario (see Figure 6). The goal of the simulation was to provide a look-and-feel of working in a human robot team (as a preparation for the values workshop one week later) and to obtain feedback on the experienced MHC within this early design phase. A number of participants remarked that they considered MHC important, but that it should not be achieved at the cost of a higher human workload (i.e. requiring the human to approve many of the robot's activities). This finding led us to define an additional requirement, about limiting communication load, for the next design cycle. One way to limit communication load while nevertheless achieving MHC would be by investing in prior control mechanisms.

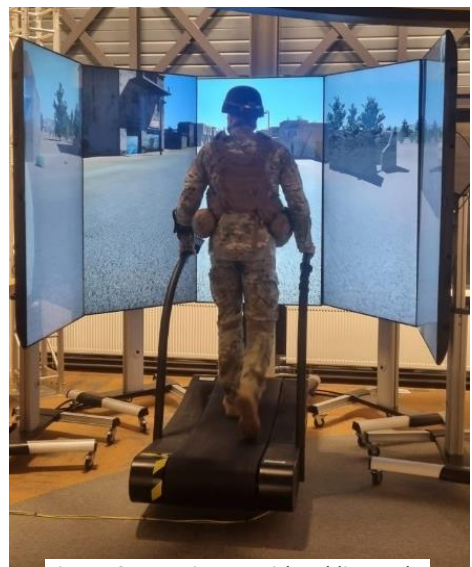

*Figure 6: Experiment with soldier and robot in a virtual environment*

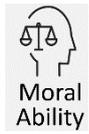 Another important outcome concerned feedback regarding Situation Awareness. Participants indicated that not knowing what the robot was doing while it was out of sight, hampered their experience of having MHC. Furthermore, participants mentioned to find identification of friend or foe difficult (i.e. not being able to ensure the robot in their field of vision was not hostile). We used this input from participants to formulate additional requirements for the next design cycle to support awareness of the robot's location and activities.

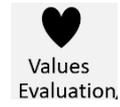 One week later, participants returned for a *Values Workshop* in the tradition of value-sensitive design (VSD) [16]. The two goals of the workshop were: to identify relevant values that military personnel have when conducting such a mission, and to discuss relationships and priorities between values with the military domain experts. The workshops used a structured walkthrough of the design scenario. The input of the participants was used to develop a value map (see Figure 8). The values are organized in different layers: at the mission level (blue), at the robot level (red), and at the interaction level (green). The value labels were assigned by the military participants.

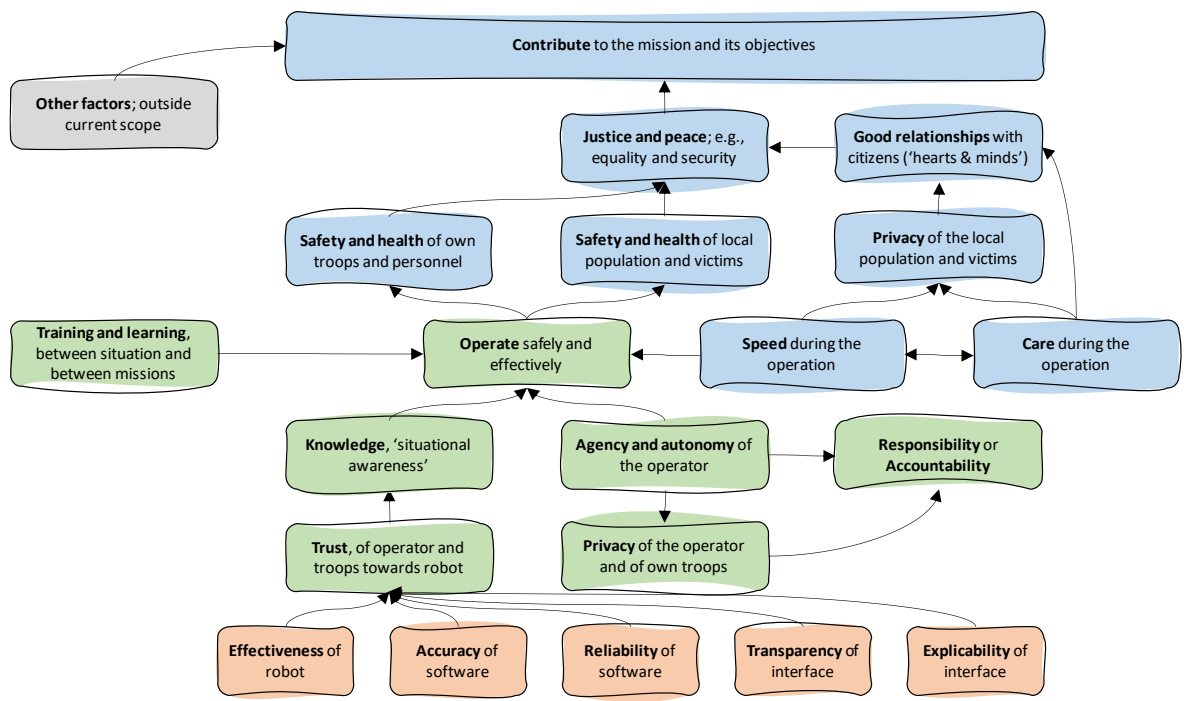

*Figure 7: Value map: mission-related values are blue; robot-related values red; and values related to HMT and MHC green*

At the mission level are values like *justice and peace*, and *safety and health*, of both citizens and troops. The team's behavior in situations that have tension between citizens' safety and own safety should be kept under MHC. One participant formulated how he balanced his priority between these tensions as follows: "If we had wanted to avoid bringing our own troops' security at risk, we might better have stayed at home". Participants also discussed how they balanced other value tensions, like e.g., maintaining *Good relationships* between citizens and troops (to care for citizens' 'hearts and minds') and respecting *privacy.* Other values discussed are related to practical operations, like *speed* and *care*, which can create a value tension, and must be therefore also be kept under MHC. One participant advocated to always put care over speed.

When participants were asked to indicate what they considered important at the robot-level, many responded with values like: *effectiveness* ("the robotic dog is only useful if it is really useful"), *accuracy and reliability (*e.g., of the image recognition software), and *transparency and explicability* (in particular with respect to the interface for aligning and communicating with the robots). Participants envisioned that the robot will have the capabilities to find and locate victims, and to perform basic observations, like e.g., monitoring the heart rate (by means of a heat camera), and detecting any injuries (e.g., by perceiving blood). Despite the fact that participants attribute advanced capabilities to future robots, they emphasize that robots should play an *auxiliary* role; people should remain responsible. Participants consider that human perception and judgement skills remain critical for triage. One potential shortcoming of robots that was identified by participants was the low visual vantage point (at 70 cm). As a result, robots may therefore easily miss victims or objects that a standing person would see (at 180 cm).

The green values in Figure 8 pertain to the operation of technology in the mission context. Participants assigned value to *trust*, referring that humans should have sufficient trust in the technology to truly contribute to the mission. Trust critically depends on the robot's functioning in terms of effectivity, accuracy, reliability, transparency and explicability. Another value that participants considered important is the robot's understanding of the situation ('situational awareness') and the robot's agency ('professional discretion'). Participants pointed out that soldiers will only rely on a robot's advice (e.g., suggesting that a victim requires immediate medical assistance, while another person is likely to be a terrorist) if they have a profound trust in the robot's technical functioning. In general, participants indicated to particularly value *knowledge* and *agency* when working with intelligent technology, and therefore considered *trust* as a critical condition for achieving MHC. When asked how they envisioned that trust in intelligent technology should develop, participants indicated that robots will likely to become better over time, and that humans who work with robots will therefore gradually develop an improved level of trust. They added, however, that trust development may be a slow and fragile process ('trust comes on foot but goes on horseback').

Participants also assigned large value to *training and learning.* Developing a collaboration with intelligent technology that is effective and efficient, while the human has and remains control over critical and moral decisions cannot be established naturally and immediately. It requires adequate training of personnel to familiarize them with the features and capabilities of the technology. Likewise, the technology too must develop some knowledge and understanding about the human team partners it collaborates with. Participants expected that some of the processes may be addressed in training situations, but that many value experiences that enable humans and agents to learn, require experiences on the job. Processes of briefing and debriefing of operations will probably be very important. Participants also assigned value to *responsibility and accountability*, which they associate with the chain of command and with legal issues concerning operations conducted by human-agent teams.

The participants' evaluation of envisioned deployment of future human-robot teams revealed many additional aspects that should be taken into account when developing a human-machine teaming concept that secures MHC, namely:
- Technology may endanger the mission: it is important to acknowledge that when an operation escalates (e.g. when people in the house need to be arrested), the robotic dog cannot do this by itself: the robot is unarmed and incapable of handcuffing humans. In addition, when the enemy detects the robot, it gives away that more military forces can be expected soon. This potential drawback, as pointed out by participants, needs to be used in the further development of the

- context design. For example, by adding a requirement: 'the robot must be co-located with humans'.
- Technology must be able to act discretely and silently: when conducting a covert military operation, a robot needs to be able to move and act silently. It is questionable whether such robots will be operational in approximately eight years from now. It may therefore appropriate to include an additional requirement into the context design, namely that robot should not be deployed in missions that require silent operations.
- Technology must be able to act flexibly according to pre-planned instructions: Participants point out that a military mission is seldom executed in full conjunction with the original plan. The unpredictability of real-world contexts often impose new challenges that demand adaptive behavior of both humans and agents. However, it may be feasible to anticipate on expected or possible deviations by means of developing details what-if scenarios, with accompanying instructions for the technology how to act accordingly. The requirement to equip robots with the capability to determine timing and execution of pre-planned instructions should be included in in the context design to warrant MHC.

# 5 Conclusion

In this chapter, we discussed design considerations for MHC in the military domain. We conclude that no *silver bullet* exists for achieving MHC over military AI-based systems. Given the precarious nature of the military, the intricacy of military decision making, and the sensitivity of moral decision making, any singular, clear-cut solution for MHC would be an oversimplification. Therefore, we need to regard MHC as a core principle guiding *all phases* of analysis, design and evaluation; as a property that is intertwined with all parts of the socio-technical system, including humans, machines, AI's, interactions, and context; and as a property that spans longer periods of time, encompassing both prior and realtime control.

Perhaps, we need to understand MHC as a new scientific discipline that has been relevant since AI-based systems have first been deployed. MHC could be like *safety research,* which has become relevant after the first technologies appeared that were dangerous for humans, such as automobiles, airplanes, and factories. Over the years, the field of safety research has specialized in several domains and provided a vast array of practical concepts such as training programs, safety officers, safety culture studies, legal norms, technology standards, supply chain regulations, etc. This analogy points out that MHC research might still be in its embryonic phase and will turn out to be even more encompassing than it currently appears.

Fortunately, such an emerging discipline does not need to start from scratch. Methods from existing design methodologies provide a head start, such as human-machine teaming, human factors, and value-sensitive design. The purpose of this chapter has been to provide a high-level design framework, and to demonstrate how different methods may be used in combination to iteratively design for MHC. Each method may potentially hold a piece of the puzzle. Designing for MHC in military human-machine teams requires a clever combination of these methods. New challenges will come up as technology advances, and HMT's become larger in the future (more interconnected humans and robots). To bring the difficult issue of MHC forward requires an ongoing international debate within academic communities, in NATO and in industry, and a watchful eye on emerging technologies and unexpected manifestations of control problems. The experiences described in this paper have shown that the iterative design approach may bring forward solutions that will establish MHC while still maintaining military operational effectivity. It is already a valuable effort to adopt MHC as a guiding principle in the design of current military technologies. Increasingly so, it will be our moral obligation to do so.

**Preprint**, to be published as: van Diggelen, J., van den Bosch, K., Neerincx, M., & Steen, M. (2023). Designing for Meaningful Human Control in Military Human-Machine Teams, in Research handbook on Meaningful Human Control of Artificial Intelligence Systems. Edward Elgar Publishing.